\documentclass[10pt,conference]{IEEEtran}
\usepackage{cite,graphicx,psfrag,amsmath,amssymb,subfigure,url}

\def\order{O}
\def\definedas{\triangleq}
\def\ATwo{{\mathop{\rm A2}}}
\def\AThree{{\mathop{\rm A3}}}
\def\R{\ }
\def\Rno{}
\def\TM{\ }
\def\TMno{}
\newcommand{\defn}[0]{\textit}

\begin{document}
\title{On Conditional Branches in Optimal Search Trees}

\author{Michael~B.~Baer \\
Electronics for Imaging\\
303 Velocity Way\\
Foster City, California 94404 USA\\
Email: Michael.Baer@efi.com}  

\maketitle

\begin{abstract} 
Algorithms for efficiently finding optimal alphabetic decision trees
--- such as the Hu-Tucker algorithm --- are well established and
commonly used.  However, such algorithms generally assume that the
cost per decision is uniform and thus independent of the outcome of
the decision.  The few algorithms without this assumption instead use
one cost if the decision outcome is ``less than'' and another cost
otherwise.  In practice, neither assumption is accurate for software
optimized for today's microprocessors, which generally have one cost
for the more likely decision outcome and a greater cost --- often far
greater --- for the less likely decision outcome.  This problem and
generalizations thereof are thus applicable to hard coding static
decision tree instances in software, e.g., for optimizing program
bottlenecks or for compiling switch statements.  An $\order(n^3)$-time
$\order(n^2)$-space dynamic programming algorithm can solve this
optimal binary decision tree problem, and this approach has many
generalizations that optimize for the behavior of processors with
predictive branch capabilities, both static and dynamic.  Solutions to
this formulation are often faster in practice than ``optimal''
decision trees as formulated in the literature.  Different search
paradigms can sometimes yield even better performance. 
\end{abstract}


\section{Introduction}

Consider a problem of assigning grades to tests.  These tests might be
administered to humans or to objects, but in either case there are
grades $1$ through $n$ --- $n$ being $5$ in most academic systems ---
and the corresponding probabilities of each grade, $p(1)$ through
$p(n)$, can be assumed to be known.  (If unknown, they are assumed to
be identical.)  Each grade is determined by taking the actual score,
$a$, dividing it by the maximum possible score, $b$, and seeing which
of $n$ distinct fixed intervals of the form $[v_{i-1}, v_i)$ the ratio
lies in, where $v_0 = -\infty$ and $v_n = +\infty$.  This process is
repeated for independently determined grades enough times that it is
worthwhile to consider the fastest manner in which to determine these
scores.

A straightforward manner of assigning scores would be to multiply (or
shift) $a$ by a constant $k$ ($\log_2 k$), divide this by $b$, and use
lookup tables on the scaled ratio.  However, division is a slow step
in most CPUs --- and not even a native operation in others --- and a
lookup table, if large, can take up valuable cache
space.  The latter problem can be solved by using a numerical
comparisons to determine the score, resulting in a decision tree.  In
fact, with this decision tree, we can eliminate division altogether; instead of
comparing scaled ratio $ka/b$ with grade cutoff value, $v_i$, we
can equivalently compare $ka$ with $bv_i$, replacing the slow division
of variable integers with a fast multiplication of a variable and a
fixed integer.  The only matter that remains is determining the
structure of the decision tree.

The desired tree has a large variety of applications --- e.g., the
compilation of switch (case) statements\cite{Sale, HeMe} --- and such
a decision tree is known as an \defn{optimal alphabetic binary tree}.
Algorithms used for finding such trees, however, find trees with
minimum expected path length, or, equivalently, minimum expected
number of comparisons\cite{Knu71,HuTu,GaWa}, whereas we want a tree
that results in minimum average run time.  The general assumption in
finding an optimal search tree is that these goals are identical, that
is, that each search step (edge) takes the same amount of time (cost)
as any other; this is noted in Section 6.2.2 of Knuth's \textit{The
Art of Computer Programming}\cite[p.~429]{Knu3}.  In exercise 33 of
Section 6.2.2, however, it is conceded that this is not strictly true;
in the first edition, the exercise asks for an algorithm for where
there is an inequity in cost between a fixed cost for a left branch
and a fixed cost for a right branch\cite{Knu31}, and, in the second
edition, a reference is given to such an algorithm\cite{Itai}.  Such
an approach has been extended to cases where each node has a possibly
different, but still fixed, asymmetry\cite{Shin}. 

In practice the asymmetry of branches in a microprocessor is different
in character from any of the aforementioned formulations.  On complex
CPUs, such those in the Pentium\R family, branches are predicted as
taken or untaken ahead of execution.  If the branch is predicted
correctly, operation continues smoothly and the branch itself takes
only the equivalent of one or two other instructions, as instructions
that would have been delayed by waiting for the branch outcome are
instead speculatively executed.  However, if the branch is improperly
predicted, a penalty for misprediction is incurred, as the results of
speculatively executed instructions must be discarded and the
processor returned to the state it was at prior to the
branch\cite{HePa3}.  In the case of the Pentium\R 4 processor, a
mispredicted branch takes the equivalent of dozens of
instructions\cite{Fog}.  This penalty has only increased with the
deeper pipelines of more recent processors.  While this time penalty
pales in comparison to that taken by division --- over a hundred adds
and shifts can take place in the time it takes to do one 32-bit
division on the Pentium\R family processors --- it certainly reveals
comparison tally as being a poor approximation to run time, the
ideal minimization.

In this paper, we discuss the construction of alphabetic binary trees
--- and more general search trees --- that are optimized with respect
to the behavior of conditional branches in microprocessors.  We
establish a general dynamic programming paradigm, one applicable to
such architecture families as: the Intel\R Pentium\R architectures,
which use advanced dynamic branch prediction; the ARM\R architectures,
most instances of which use static branch prediction; and Knuth's MMIX
architecture, in which branches explicitly ``hint'' whether or not
they are assumed taken or untaken\cite[p.~20]{Knu05}.  The first two
are not only representative of two styles of branching; they are also
by far the most popular processors for 32-bit personal computers and
32-bit embedded applications, respectively.  Dynamic programming
algorithms with $\order(n^3)$-time $\order(n^2)$-space performance are
developed and discussed.  Although these algorithms are not simplified
in the same manner as Knuth's dynamic programming
algorithm\cite{Knu71}, they are fairly flexible, accounting for
different costs (run times) for different comparisons due to such
behaviors as dynamic branch prediction and
conditional instructions other than branches.

\section{No prediction and static prediction}
\label{static}

Consider Knuth's pedagogical MMIX architecture\cite{Knu05}, which has
a simple rule for branching: If we know ahead of time which branch is
more likely and which less likely, we can hard code the more likely
branch to take $1+c$ clock cycles and the less likely branch to take
$3+c$ clock cycles, where $c$ represents the time taken by
instructions other than the branch itself, e.g., multiplications,
additions, comparisons.  Note that the disparity in performance
between correctly anticipated branches and incorrectly anticipated
branches is not as great as that for recent versions of the Pentium\R
architectures, in which the exact number depends on the processor
architecture and the type of comparison.  For this ubiquitous family
of processors, properly modeling the asymmetry of search tree
performance is even more vital than it is on MMIX.  However, since the
MMIX architecture uses hints with no branch prediction, it is simpler
to model and to hard code any desired preference.  (Some real-world
processors such as the Intel\R Pentium\R 4\cite[p.~2-2]{IntP} and the
MIPS R4000\cite[p.~21]{Hein} also use such branch hints, usually in
conjunction with more advanced prediction techniques discussed in the
next section.)

It is also easy to code the asymmetric bias of the branch for most
implementations of static branch prediction.  In static prediction,
opcode or branch direction is used to determine whether or not a
branch is taken, the most common rule being that forward conditional
branches are presumed taken and backward conditional branches are
presumed not taken\cite{HePa3}.  Assume, for example, that we want to
use a forward branch, which is assumed not to be taken.  We thus want
the most likely outcome to be that the branch is not taken: For
example, if it is more likely than not that the item is less than $v_i$, the
branch instruction should correspond to ``branch if greater than or
equal to $v_i$.''  Otherwise, the branch instruction should correspond
to ``branch if less than $v_i$.''

\begin{figure*}[ht]
     \centering
     \psfrag{1}{$1$}
     \psfrag{3}{$3$}
     \subfigure[Fixed-length representation]{
          \label{fixedtree} 
          \includegraphics{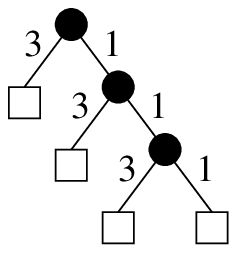}}
     \quad
     \subfigure[Proportional representation]{ 
          \label{proptree}
          \includegraphics{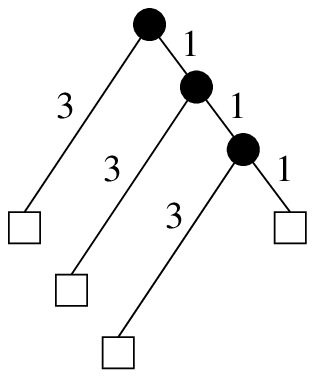} }
     \caption{Two representations of optimal branch trees}
     \label{branchtrees}
\end{figure*}

The decision tree problem, applicable to problems with either no
branch prediction or static branch prediction, considers positive
weights $c_1$ and $c_2$ such that the cost of a binary path with
predictability $b_1b_2 \cdots b_k$ is
$$t(b_1b_2 \cdots b_k) \definedas \sum_{j=1}^k c_{b_j}$$ where $b_j =
1$ for a mispredicted result and $b_j = 2$ for a properly predicted
result.  Such tree paths are often pictorially illustrated via longer
edges on the corresponding tree, so that path height corresponds to
path cost, e.g., Fig.~\ref{proptree}.  Thus the overall expected cost
(time) to minimize is
\begin{equation}
T(b) \definedas \sum_{i=1}^n p(i) \sum_{j=1}^{l(i)} c_{b_j(i)}
\end{equation}
where $p(i)$ is the probability of the $i$th item, $l(i)$ is the
number of comparisons needed, and $b_j(i)$ is $1$ if the result of the
$j$th branch for item $i$ is contrary to the prediction and $2$ otherwise.

Note that if the number of comparisons is the value to be emphasized
in pictorial representation, edges can be portrayed with fixed height,
as in Fig.~\ref{fixedtree}.  In Fig.~\ref{proptree}, by contrast, the
total cost is the value emphasized; having edges portrayed in this
way, with height proportional to their cost, is usually preferred.

This problem can be placed in the context of optimal binary tree
problems, as in Table~\ref{types}.  Other than the \textit{branching
problem} considered in this paper, problems are referred to as in
\cite{Abr01}.  In most problem formulations, edge cost is fixed, and,
where it is not fixed, edges generally have costs according to their
order, i.e., a left edge has cost $c_1$ and a right edge has
cost $c_2$.  Relaxing this edge order constraint in the unequal cost
alphabetic problem results in the problem we are now considering.
Note that Karp's nonalphabetic variant does not change if edge order
is allowed to change; since output items need not be in a given (e.g.,
alphabetical) order, the tree optimal for the ordered-edge
nonalphabetic problem is also optimal for the unordered-edge
nonalphabetic problem.  Because of all this, the cost for the optimal
tree under Karp's formulation is a lower bound on the cost of the
optimal branch tree, whereas the cost for the optimal tree under
Itai's formulation is an upper bound on the cost of the optimal
branch tree.  This enables the use of the bounds formulated in
\cite{AlMe} for the branching problem.  

\begin{table*}
\centering
\begin{tabular}{l||c|c|c}
&\multicolumn{3}{c}{\textit{Edge cost/order specification}} \\[6pt]
\textit{Output order restriction} &\textbf{Constant cost} &\textbf{Order restricted} &\textbf{Order unrestricted} \\[6pt]
\hline
\hline
&&& \\[-4pt]
\textbf{Alphabetic}    &Hu-Tucker\cite{GiMo,HuTu,GaWa,Knu3} 
  & Itai\cite{Itai,Shin} & branching problem \\[6pt]
\hline
&& \multicolumn{2}{c}{} \\[-4pt]
\textbf{Nonalphabetic} &Huffman\cite{Huff,Leeu,Knu}
  & \multicolumn{2}{c}{Karp\cite{Karp,GoRo,BGLR}} \\[8pt]
\end{tabular}
\caption{Types of decision tree problems}
\label{types}
\end{table*}

The key to constructing an algorithm is to note that any optimal
alphabetic tree must have all its subtrees optimal; otherwise one
could substitute an optimal subtree for a suboptimal subtree,
resulting in a strict improvement in the result.  Each tree can be
defined by its \defn{splitting points}.  A splitting point $s$ for the
root of the tree means that all items after $s$ and including $s$ will
be in the right subtree while all items before $s$ will be in the left
subtree.  Since there are $n-1$ possible splitting points for the
root, if we know all potential optimal subtrees for all possible
ranges, the splitting point can be found through sequential search of
the possible combinations.  The optimal tree is thus found
inductively, and this algorithm has $\order(n^3)$ time complexity and
$\order(n^2)$ space complexity, in a similar manner to \cite{GiMo}.

The dynamic programming algorithm, given the aforementioned
considerations, is relatively straightforward.  Each possible optimal
subtree for items $i$ through $j$ has an associated cost, $c(i,j)$ and
an associated probability $p(i,j)$; at the end, $p(1,n)=1$ and
$c(1,n)$ is the expected cost (run time) of the optimal tree.  


The base case and recurrence relation we use are similar to those of
\cite{Itai}.  Given unequal branch costs $c_1$ and $c_2$ and
probability mass function $p(\cdot)$ for $1$ through~$n$,
\begin{equation}
\begin{array}{rcl}
c(i,i) &=& 0 \\ 
c'(i,j) &=& \min_{s \in (i,j]} \{c_1 p(i,s-1) + c_2 p(s,j) +{} \\
&& \quad c(i,s-1) + c(s,j)\} \\
c''(i,j) &=& \min_{s \in (i,j]} \{c_2 p(i,s-1) + c_1 p(s,j) +{} \\
&& \quad c(i,s-1) + c(s,j)\} \\
c(i,j) &=& \min\left\{c'(i,j),c''(i,j)\right\}
\end{array}
\label{opt}
\end{equation}
where $p(i,j) = \sum_{k=i}^j p(i)$ can be calculated on the fly along
with~$c(i,j)$.  The last minimization determines which branch
condition to use (e.g., $<$ vs. $\geq$ or ``assume taken''
vs. ``assume untaken''), while the minimizing value of $s$ is the
splitting point for that subtree.  The branch condition to use ---
i.e., the bias of the branch --- must be coded explicitly or
implicitly in the software derived from the tree.

Knuth\cite{Knu71} and Itai\cite{Itai} also begin with such an
algorithm, then go on to note that the splitting point of an optimal
tree for their problems must be between the splitting points of the
two (possible) optimal subtrees of size $n'-1$, and use this fact to
reduce complexity.  The branching problem considered here, however,
lacks this property.  Consider $p = (0.3~0.2~0.2~0.3)$ and $c =
(3~1)$, for which optimal trees split either at $2$, as in
Fig.~\ref{branchtrees}, or at $4$, the mirror image of this tree.  In
contrast, the two largest subtress, as illustrated in the figure and
its mirror image, both have optimal split points at~$3$.  Similarly,
applying the less complex Hu-Tucker approach\cite{HuTu,HKT} to this
problem fails for $p = (0.2~0.15~0.15~0.2~0.3)$ and $c = (3~1)$.

\begin{figure*}[ht]
     \centering
     \subfigure[Edge order restricted]{ 
          \resizebox{6cm}{!}{\includegraphics{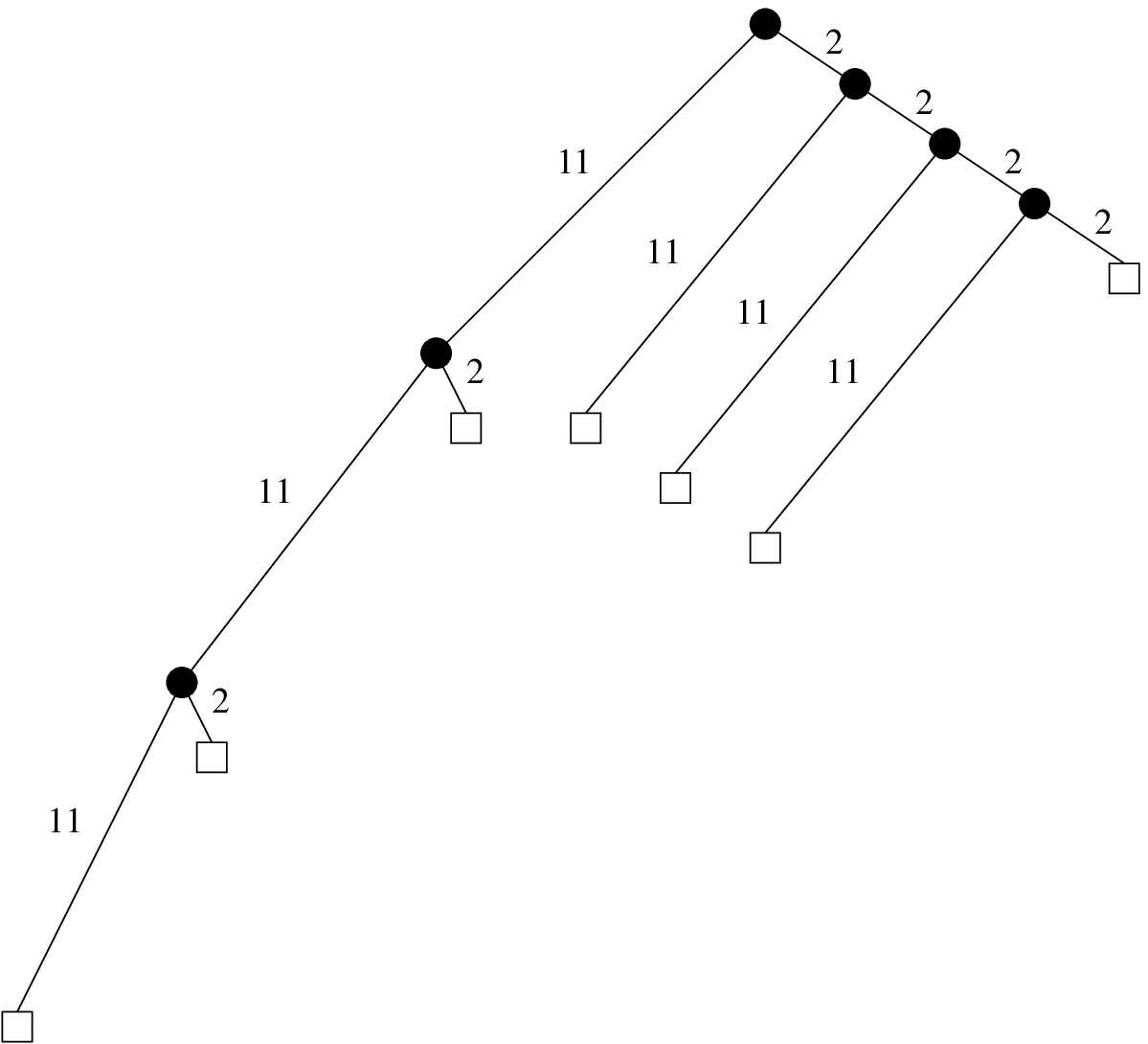}}
          \label{bitai}}
     \subfigure[Edge order unrestricted]{
          \resizebox{7.75cm}{!}{\includegraphics{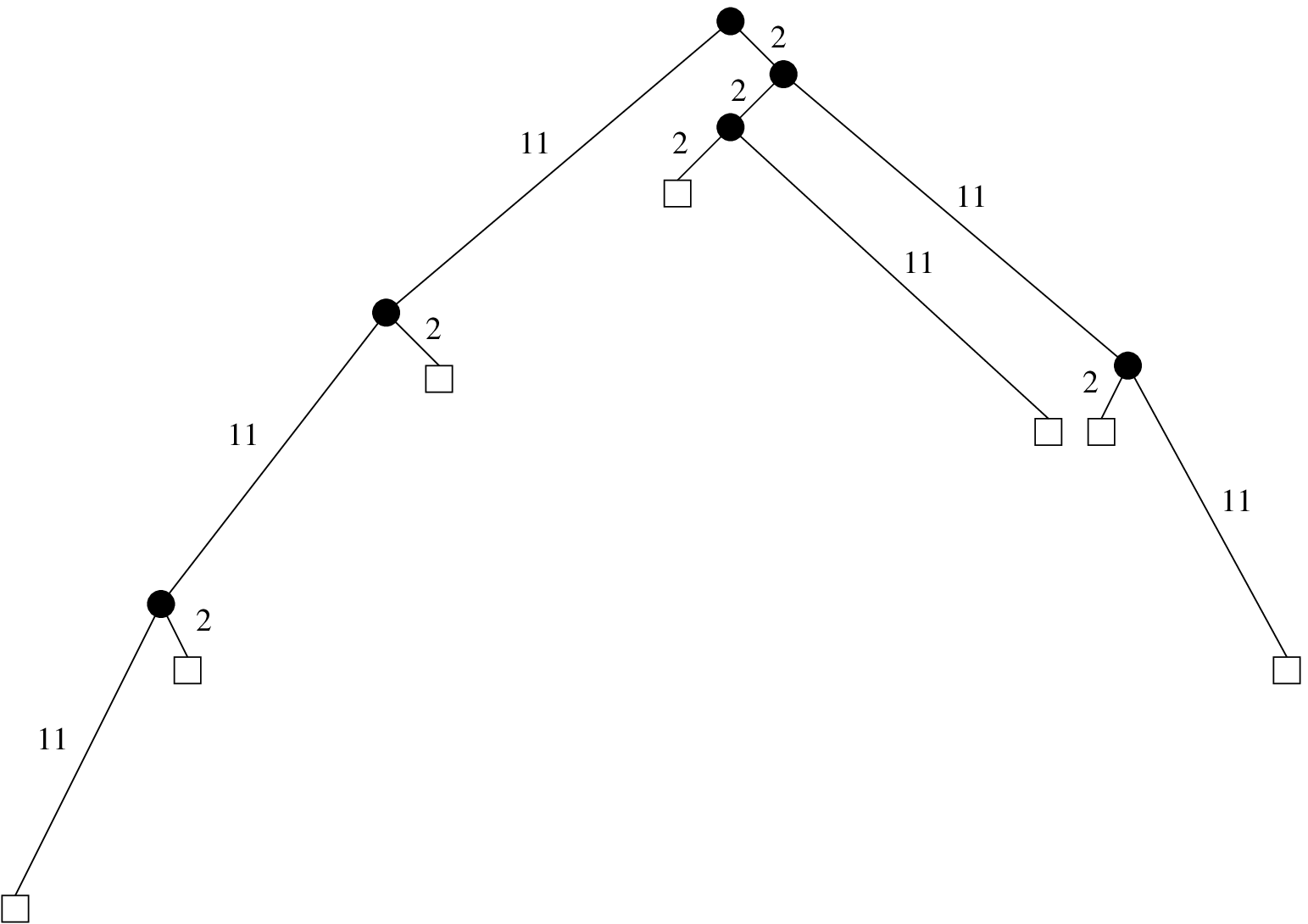}}
          \label{bbr}}
     \caption{Optimal branch trees for two restriction types for $p = (1\ 6\ 15\ 20\ 15\ 6\ 1)/128$ with $c = (11~2)$}
     \label{orders}
\end{figure*}

The optimal tree of Fig.~\ref{branchtrees} is identical to the optimal
tree returned by Itai's algorithm for order-restricted
edges\cite{Itai}.  Consider a larger example in which this is not so,
the binomial distribution $p = (1\ 6\ 15\ 20\ 15\ 6\ 1)/128$ with $c =
(11~2)$.  If edge order is restricted as in \cite{Itai}, the tree at
Fig.~\ref{bitai} is optimal, yielding an expected cost of $967/64 =
15.109375$.  If we relax the restriction, as in the problem under
consideration here, the tree at Fig.~\ref{bbr} is optimal, yielding an
expected cost of $831/64 = 12.984375$, a $14\%$ improvement.

The $\order(n^3)$ complexity used to get this better result is
generally not limiting.  Although this restricts the range of problems
solvable within a given time, most hard-coded search tree problems are
small enough to be solved quickly.  In addition, this model is not
robust enough to model performance for large seach trees, where some
problems, such as the aforementioned grade-assignment problem, are
actually better solved via alternative means when $n$ is large; a
division and lookup table might be more suitable, for example, if
there are dozens of different grades.  This is not only due to the
fact that search run time scales as $\order(\log n)$, but also because
large blocks of software code generally run more slowly due to
insufficient instruction caching.  If larger problems did need to be
solved within this framework, space complexity would likely become an
issue before time complexity anyway.

Given no information on $p(\cdot)$, it is common to assume that
$p(i)=1/n$.  This is justified by noting that, if $p(\cdot)$ is
considered a random vector drawn from the probability simplex
according to density $g$ --- as in \cite{Cov72} --- the expected cost for a given coding scheme
is
$$E_p[T(b)] = E_p\left[\sum_{i=1}^n p(i) \sum_{j=1}^{l(i)}
c_{b_j(i)}\right] = \sum_{i=1}^n E_p[p(i)] \sum_{j=1}^{l(i)}
c_{b_j(i)}$$ and, if $g$ is symmetric on the
simplex, $E[p(i)] = 1/n$ for any~$i$.  Since edge costs are not
fixed, optimal trees for this uniform distribution need not be
complete trees, that is, full trees for which all leaves have depth
$\lfloor \log_2 n \rfloor$ or $\lceil \log_2 n \rceil$.  For example,
the tree in Fig.~\ref{branchtrees} is optimal for this uniform
distribution with an average cost of $3.75$, whereas the complete tree
for $n=4$ results in an average cost of $4$.  This is thus a better
approach to use for compilers that code switch (case) statements
partially\cite{HeMe} or entirely\cite{Sale} as decision trees. 

ARM\R architectures such as those of the ARM7\TM and ARM9\TM families
use no or static branch prediction\cite{ArmPC}.  Such processors are
used for most mobile devices, including cell phones and iPods\Rno.
Recent Pentium\R designs and the XScale\TMno \cite{IntX} --- which is
viewed as the successor to ARM\R architecture StrongARM\R --- use
dynamic prediction, which we explore in the next section.

\section{More advanced models}

With dynamic branch prediction, which in more advanced forms includes
branch correlation, branches are predicted based on the results of
prior instances of the same and different branch instructions.  This
results in improved branch predictability for most software
implementations, especially those in which branch profiling does not
enter into software design.  Where branch profiling does take place,
however, the gains are often only marginal\cite[pp.~245--248]{HePa3}.
Thus, although large processors and general purpose processors
generally include dynamic prediction, many small processors and
low-power processors forgo dynamic prediction, as this feature's
sophistication requires the usage of significant additional
semiconductor area for the associated logic.

Dynamic branch prediction often results in complex processor behavior.
Often several predictors will be used for the same branch instruction
instance; the predictor in a given iteration will be based on the
history of that branch instruction instance and/or other branches.  In
the problem we are concerned with, however, this does not result in as
many complications as one might expect; the probability of a given
branch outcome conditional on the branches that precede it is
identical to the probability of the branch outcome overall.  In the
case of previous branch outcomes for the same search instance ---
i.e., those of ancestors in the tree --- any given outcome is
conditioned on the same events --- i.e., the events that lead to the
branch being considered.  In the case of branches for previous items,
if items are independent, so are these branches.  In the case of
branches outside of the algorithm, these can also be assumed to be
either fixed given or independent of the current branch.

Thus, as long as each branch predictor is assigned at most one of the
decision tree branches, prediction can be modeled as a random process.
This process will result in each predictor converging to a stationary
distribution, which can be analyzed and optimized for.  Unfortunately,
such a random process will necessarily perform worse than optimized
static prediction, although, in most instances, the difference will
not be too great.

The cost of each branch result can be determined by the expected
time of the branch, based on the costs involved and the probability
that the branch is correctly predicted.  Simple analysis of the
stationary distribution of a branch prediction Markov chain, e.g.,
\cite{HPS}, can yield the expected time for a given branch direction
as a function of the probability of the branch.  

\begin{figure}[ht]
     \centering 
     \psfrag{N}{{\small N}}
     \psfrag{T}{{\small T}}
     \psfrag{0/N}{{$0$/N}}
     \psfrag{1/N}{{$1$/N}}
     \psfrag{2/T}{{$2$/T}}
     \psfrag{3/T}{{$3$/T}}
     \subfigure[A2 (recent Pentium\R architectures)]{
          \includegraphics{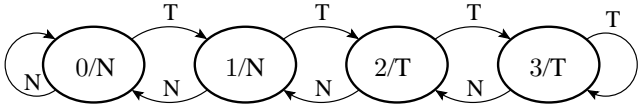}
          \label{A2}}
     \qquad
     \subfigure[A3 (\textit{Computer Architecture -- A Quantitative Approach})]{ 
          \includegraphics{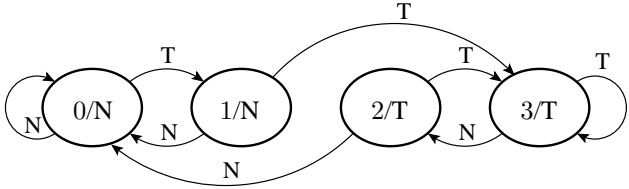}
          \label{A3}}
     \caption{Moore state diagrams for branch prediction}
     \label{state}
\end{figure}

For example, if branch prediction uses a saturating up-down counter
--- the two-bit Markov chain of \cite{PSR} shown as a Moore state
diagram in Fig.~\ref{A2} --- then the probability of misprediction is
given by
$$f_{\ATwo}(p_1) \definedas P[\mbox{mispredict on A2}] =
\frac{p_1-p_1^2}{1-2p_1+2p_1^2}$$ where $p_1$ is the probability the
less likely event will occur given the branch being considered.  This
Markov chain is used by the more recent Pentium\R
architectures\cite{Fog} and is  
referred to by Yeh and Patt as Automation A2\cite{YePa}.  
If branch prediction instead uses the two-bit Markov
chain of \cite{McHe}, as in the MIPS-influenced pedagogical
architecture in \cite{HePa3} and in Fig.~\ref{A3}, then the
probability of a misprediction is given by
$$f_{\AThree}(p_1) \definedas P[\mbox{mispredict on A3}] =
\frac{p_1+p_1^2+4p_1^3+2p_1^4}{1-p_1+p_1^2}.$$ This chain is referred
to by Yeh and Patt as Automation~A3.  (Other state diagrams considered
by Yeh and Patt are not in as wide use, either being too simple or
lacking symmetry between taken and untaken branches.)

\begin{figure}[t]
     \centering
     \resizebox{8.5cm}{!}{\includegraphics{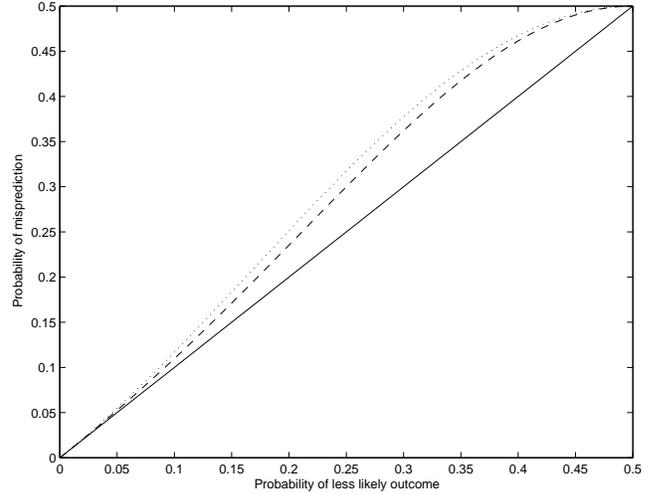}}
     \caption{Static and dynamic branch misprediction rates}
     \label{dynmis}
\end{figure}

The Moore state diagrams for Automations A2 and A3 in Fig.~\ref{state}
have N representing an untaken branch and T representing a taken
branch.  The state determines which prediction is made, and the actual
branch outcome determines the transitions between the states.  The
corresponding stationary misprediction rate (Fig.~\ref{dynmis}) is worst for
Automation A3 (dotted plot), which has a rate up to 26.92\% worse than
that for static prediction (solid plot).  Automatic A2 (dashed plot)
does better, being at most 20.71\% worse than static prediction.

In either dynamic case, we can calculate expected cost from the
probabilities of each branch.  We initially assume that taken and
untaken branches are symmetric, that is, a branch rightly predicted as
untaken will have the same cost as a branch rightly predicted as
taken, and mispredicted branches will similarly have identical costs.
Let the type of dynamic prediction be $A$, let the probability of the more
likely subtree be $p_{\max}$, and let the probability of the less likely
subtree be $p_{\min}$, so that $p_{\min} + p_{\max} \leq 1$ and $p_1 =
p_{\min}/(p_{\min} + p_{\max})$ is the probability of the less likely
outcome conditional on the branch being decided upon.  Then, instead
of the static cost of $c_1 p_{\min} + c_2 p_{\max}$, the expected cost
of a given branch is
$$
\begin{array}{l}
C(p_{\min},p_{\max}) \definedas \\[4pt]
\quad c_1 (p_{\min}+p_{\max}) f_A\left(\frac{p_{\min}}{p_{\min}+p_{\max}}\right) +{} \\[6pt] 
\qquad c_2 (p_{\min}+p_{\max})\left(1-f_A\left(\frac{p_{\min}}{p_{\min}+p_{\max}}\right)\right).
\end{array}
$$
Thus this plus the costs of the subtrees is the overall tree cost.

Asymmetries in taken and untaken branches are easily accounted for.
Similarly, if a $(<,\geq)$ comparison with a certain value has a
smaller cost than a comparison with another value --- say a comparison
with a power of two times a variable is faster due to the inherently
reduced calculation time --- then this can also be taken into account.
Another variant can be found by noting that some processors allow
conditional instructions, that is, instructions only executed given
certain conditions.  On platforms such as those in the Pentium\R
family, a conditional instruction is often preferable to a conditional
branch, but this might only reasonably be used to eliminate a branch
to leaves in the decision tree.  Thus branches deciding between only
two items might be accounted differently.  

With such a variety of options, there could be multiple possible costs
for any particular branch.  General cost functions for a branch taking
all this into account are of the form $C_k(p',p'',i,j,s)$ for $k$ from
$1$ to some $m$ (where $m=2$ is most common), and (\ref{opt}) thus becomes:
\begin{equation}
\begin{array}{rcl}
\displaystyle
c(i,i) &=& 0 \\[4pt]
c_k(i,j) &=& \displaystyle
\min_{s \in (i,j]} \{C_k(p(i,s-1),p(s,j),i,j,s) +{}\\
&& \quad c(i,s-1) + c(s,j)\}  \quad \forall k \\[4pt]
c(i,j) &=& \displaystyle \min_{k \in [1,m]} \left\{c_k(i,j)\right\}
\end{array}
\label{general}
\end{equation}
Once again, this is a simple matter of dynamic programming, and,
assuming all $C_k$ are calculable in constant time, this can be done
in $\order(mn^3)$ time and $\order(n^2 + n \log m)$ space, the $\log
m$ term accounting for recalculation and storage of the type of cost
function (decision method) used for each branch.  An even more general
version of this could take into account properties of subtrees other
than those already mentioned, but we need not consider this here.

Note that, because dynamic prediction is adaptive to dynamic branch
performance, we need not explicitly code branch bias; the more and
less likely branch outcomes will automatically be detected.  However,
most dynamic predictors begin with state that depends on the type of
branch in the same manner as static prediction.  That is, the first
time a branch is encountered, it is usually statically predicted.
Thus it might also be worthwhile to implement the search software so
that initial iterations of the search tree behave as well as possible
given the tree optimal for asymptotic behavior.  Note that this
variation adds no time or space complexity to the above
algorithm.

If the software is predetermined but the tree is not --- that is, if
the software is not hard coded for the specific tree --- then matters
change entirely; no prediction and static prediction result in this
problem being equivalent to that of ordered edges, the problem
proposed by Knuth and considered by Itai.  Dynamic prediction with
correlations, on the other hand, can result in a number of outcomes,
depending on implementation.  The software on a given processor could
have near-perfect distinguishing of outcomes, in which the above
dynamic analysis persists.  More likely, without sufficient
unrolling\cite{HePa3} of the tree data structure, there would be
confusion of outcomes, as the software would not know whether all
previous branches untaken indicates we are at the root, at its right
child, at its right child's right child, etc.  Optimizing for the
complex dependencies involved with such a system is no longer within
the above framework, and the overall averaging effect means that one
might want to just use the tree optimal for the corresponding static
problem Therefore the aforementioned methods are usually best suited
for when the user has the option of designing software specifically
for a given decision tree, or designing hardware and/or programmable
logic to allow the methods to work in fixed software.

\section{Search trees and equality comparisons}

Knuth showed how a dynamic programming approach can be used for
general search trees\cite{Knu71}, in which the decision is no longer
binary, but is instead, ``Is the output greater than, less than, or
equal to $x$?''  This allows items to be implicitly or explicitly
stored within the internal nodes (nonleaves) of the decision tree and
allows us to consider items that might not be in a search tree.  This
model generalizes the concept of an alphabetic decision tree and can
be used for applications in which there is an inherent ``dictionary''
of items, such as token parsing and spell checking.  Probabilities for
both present and missing items are then needed.

Before formalizing this, we should note a few things about the
applicability of the search tree model.  Clearly the problem at the
beginning of this paper does not fall into this model, as strict
equality cannot be tested for.  Even where this model is applicable,
it can be too restrictive.  For example, this model is often
inferior to the alphabetic model for the simple reason that, on most
hardware, including all hardware considered here, three-way branches
are not native operations.  They must thus be simulated by two two-way
branches in a manner that actually results in greater run time.
Experimental analysis of this phenomenon can be found in \cite{Ande}
and numerical analysis can be found in \cite{HuTu2} and
\cite[pp.~344--345]{HuSh}.  These all find that an alphabetic tree is
usually preferable in practice.  Thus we only briefly discuss issues
of this search tree model.

\begin{figure*}[ht]
     \centering
     \subfigure[Alphabetic tree (${\beta}_i=0$)]{ 
          \label{atree}
               \psfrag{a0}{$p(1)$}
               \psfrag{a1}{$p(2)$}
               \psfrag{a2}{$p(3)$}
               \psfrag{a3}{$p(4)$}
               \psfrag{a4}{$p(5)$}
          \includegraphics{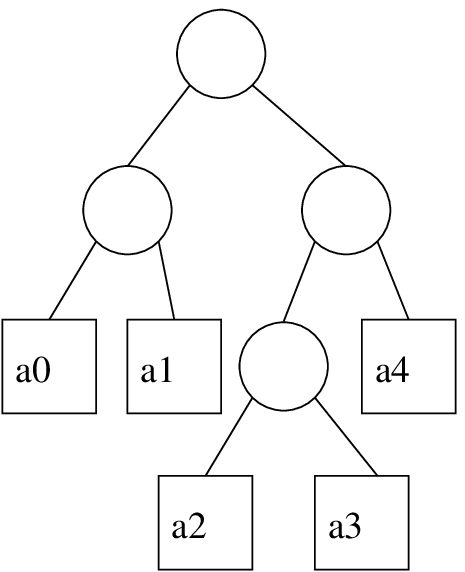} }
     \subfigure[Search tree (${\alpha}_i=0$)]{
          \label{btree} 
               \psfrag{b1}{$p(1)$}
               \psfrag{b2}{$p(2)$}
               \psfrag{b3}{$p(3)$}
               \psfrag{b4}{$p(4)$}
          \includegraphics{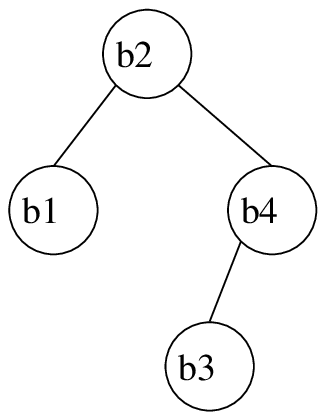} }
     \subfigure[Full search tree]{ 
          \label{stree} 
               \psfrag{b1}{$\beta_1$}
               \psfrag{b2}{$\beta_2$}
               \psfrag{b3}{$\beta_3$}
               \psfrag{b4}{$\beta_4$}
               \psfrag{a0}{$\alpha_0$}
               \psfrag{a1}{$\alpha_1$}
               \psfrag{a2}{$\alpha_2$}
               \psfrag{a3}{$\alpha_3$}
               \psfrag{a4}{$\alpha_4$}
          \includegraphics{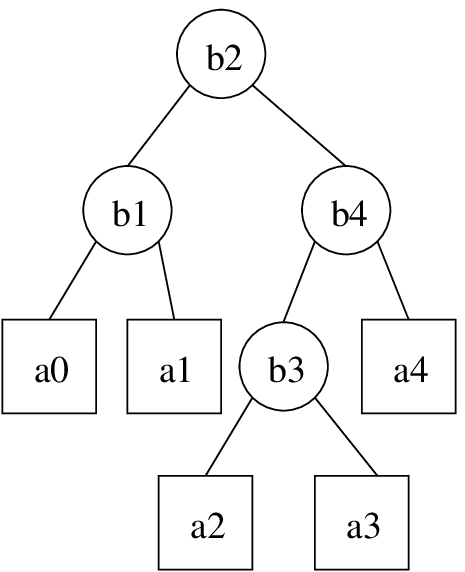} }
     \caption{Sample search trees (fixed-length representation)}
     \label{searchtrees}
\end{figure*}

For items $1$ through $n'$, $\beta_i$ is defined as the probability
that a search yields item $i$ and $\alpha_i$ as the probability that a
search fails and the item not in the search tree would be
lexicographically between items $i$ and $i+1$; $\alpha_0$
is the probability it is before $\beta_1$ and $\alpha_n$ is the
probability it is after~$\beta_n$.  Thus
$$\sum_{i=1}^{n'} \beta_i + \sum_{i=0}^{n'} \alpha_i = 1 .$$ The
alphabetic tree scenario is a special case, with $n'=n-1$,
$\beta_i=0$, and $\alpha_i = p(i+1)$, as in Fig.~\ref{atree} for
$n=5$.  Fig.~\ref{btree} is a similar search tree configured for a
three-way comparison; this time, there are only four items, and it is
assumed that all items searched for will be in the tree.
Fig.~\ref{stree} is the same four-item search tree allowing one to
search for both items in the tree (with probabilities $\{\beta_i\}$)
and ranges of missing items (with probabilities $\{\alpha_i\}$).

In such a model, there are now three costs associated with a given
node; the cost of the two branches $c_1$ and $c_2$, and the cost of an
equality, $e$.  Thus, for static prediction or no prediction,
(\ref{opt}) becomes:
\begin{equation*}
\begin{array}{rcl}
c(i,i) &=& 0 \\ 
p(i,i) &=& \alpha_i \\
c'(i,j) &=& \min_{s \in (i,j]} \{c_1 p(i,s-1) + c_2 p(s,j)+{} \\
&& \quad e \beta_s + c(i,s-1) + c(s,j)\} \\
c''(i,j) &=& \min_{s \in (i,j]} \{c_2 p(i,s-1) + c_1 p(s,j)+{} \\
&& \quad e \beta_s + c(i,s-1) + c(s,j)\} \\
c(i,j) &=& \min\left\{c'(i,j),c''(i,j)\right\} \\
p(i,j) &=& p(i,s-1) + \beta_s + p(s,j) \quad \forall s
\end{array}
\end{equation*}
where the root case is $c(0,n')$ and $p(i,j)$ is usually calculated
using the optimizing $s$ for the overall subtree in question.  

Again, one can generalize this as in (\ref{general}); for example, the
cost of an equality comparison need not be fixed.  A further
generalization in which the equality comparison key value is different
than the inequality comparison value has been considered for the
constant edge cost version of this problem, e.g., \cite{HuWo2,HHHW}.
Approaches for solving this have led to the more germane problem in
which, rather than allowing an inequality \textit{and} an equality
comparison in each step, one allows an inequality \textit{or} an
equality comparison in each step.  This is known as the \defn{two-way
key comparison} problem.  If data are highly irregular such that the
most probable item is much more probable than any other and is in
$[2,n-1]$, then an initial equality comparison to the most probable
item would likely improve on the ``optimal'' $(<,\geq)$ decision tree.
For fixed edge costs, the algorithm for solving this is a
$\order(n^5)$-time $\order(n^3)$-space dynamic programming
algorithm\cite{Spul}.  In this algorithm, instead of just $i$ and $j$,
a third variable $d$ represents the number of items missing from the
subtree due to equality comparisons above this subtree; this accounts
for the increased complexity.  This algorithm uses the conjecture that
equality comparisons should always be with the most likely (remaining)
item.  This was not proved for equal edge costs, and, even given its
veracity, it is not clear whether this would also be true for unequal
edge costs.  Nevertheless, no counterexample has been presented, so it
is a safe assumption to make, especially since such trees would
necessarily perform at least as well as the optimal binary decision
tree.

The two-way comparison algorithm has been extended to a large variety
of problems, including a problem with nine different branch costs:
unequal (ordered) costs for \hbox{$(=,\neq)$} testing, unequal
(ordered) costs for \hbox{$(\leq,>)$} testing, unequal (ordered) costs
for \hbox{$(<,\geq)$} testing, and unequal (ordered) costs for
three-way testing\cite[Chapter~9]{Spul0}.  This algorithm can be
easily modified for unordered costs by adding tests for
\hbox{$(\neq,=)$}, \hbox{$(>,\leq)$}, \hbox{$(\geq,<)$}, and other
three way tests.  Other modifications can be made in a similar manner
to those discussed in this paper.  Note that some variants of this
problems have complexity reduced from $\order(n^5)$ to
$\order(n^4)$\cite{Spul,AKKL}, although this has not been shown to be
true of the more general cases that most accurately represent the
behavior of hard-coded search trees.

\section{Conclusion}

In this paper, we presented methods for finding optimal decision and
search trees given the real-world behavior of microprocessors, in
which not all queries and decision outcomes have identical temporal
costs.  This approach most often assumes we can hard code the decision
tree based on a known probability distribution and known processor
behavior.  The simplest method, that of Section~\ref{static}, must be
generalized for more complex processor prediction techniques, as well
as for other subtler performance considerations and for cases in which
equality comparisons are allowed.  Due to the growing asymmetry of
branch performance in complex processors, this often results in
strictly better hard-coded search trees than the ``optimal'' trees
produced using traditional methods.

\end{document}